\documentclass[12pt]{article}

\usepackage{amsmath,amssymb,amsfonts,graphicx}
\usepackage[mathscr]{euscript}
\DeclareSymbolFont{rsfs}{U}{rsfs}{m}{n}
\DeclareSymbolFontAlphabet{\mathscrsfs}{rsfs}

\numberwithin{equation}{section}
\begin{document}

\title{Euclidean quantum gravity for Kerr-Newman Spacetimes}
\author{Rohan Raha}
\date{}
\maketitle
% inline

\centerline{\Large Abstract}
It is very important to calculate the amount of radiation from a black hole as the radiation from a black hole contributes to its entropy. In this paper I have calculated the Entropy of a black hole from the action. I have used path integral formalism to calculate the second order perturbation of the metric in the action for a generalized Kerr-Newman metric.

\section{Introduction}
The "no hair theorem" states that a black hole formed in a gravitational field will collapse rapidly in a quasistationary state which can be completely characterized by three parameters:mass M, Charge Q, and angular momentum J. However,classically there will be multiple internal configurations corresponding to a particular set of values of the three parameters. According to classical theory the number of possible configurations will be infinite but taking quantum mechanics into account the number of possible states will be restricted by the fact that the wavelength of the particle should be less than the size of the black hole formed from gravitational collapse. Thus, there is supposed to be a finite number of configurations for a black hole which will give rise to a finite entropy and the black hole will emit radiation like a body with temperature as given by S.Hawking
\begin{equation}
T_H=G^2[(\frac{\partial S}{\partial M})_{J,Q}]^{-1}
\end{equation}
where $T_H$ is the characteristic temperature,S is the entropy of the black hole, M is the mass, J is the angular momentum and Q is the charge of the black hole.
The aim of this paper is to calculate the entropy for a Kerr-Newman black hole using the path integral formalism upto second order perturbation and to evaluate the partition function from which we can calculate the partition function to evaluate the entropy of a black hole with mass M, charge Q and angular momentum J. Path integral formalism is preferred in quantization of gravity as higher dimensions in String theory introduces background dependence of spacetime while cannonical quantization and loop quantum gravity breaks general covariance of spacetime.  
The partition function using the path integral approach is given by
\begin{equation} 
Z=\int d[g]d[\phi]exp\{iI[g,\phi]\} 
\end{equation}
where $I[g,\phi]$ is the action, $d[g]$ is integration over the space of metric  and $d[\phi]$ is the path integral measure over the space of all matter fields.Gibbons and Hawking has calculated the action for a Schwarzchild black hole without charge or angular momentum; here we will adopt a similar methodology to calculate the classical contribution to the action i.e. the zeroth order action for a generalized Kerr Newman black hole.
In the second part of the paper we will see how to calculate the quantum partition function by introducing perturbations to the the background metric.

\section{The Zeroth order contribution to the action}
The background solution here is taken to be the Kerr-Newman spacetimes.The kerr-newman line element in B-L coordinate with God's unit $8\pi G=\hbar=1=c=\frac{1}{4\pi\epsilon_{0}}$  is
\begin{equation}
ds^{2}=-(\frac{dr^{2}}{\Delta}+d\theta^{2})\rho^{2}+(dt-a\sin^{2}\theta d\phi)^{2}\frac{\Delta}{\rho^{2}}-((r^{2}+a^{2})d\phi-adt)^{2}\frac{\sin^{2}\theta}{\rho^{2}} 
\end{equation}
where 
\begin{equation}
\begin{split}
a=\frac{J}{M},\\
\Delta=r^{2}-r_{s}r+a^{2}+r_{Q}^{2},\\
\rho^{2}=r^{2}+a^{2}\cos^{2}\theta,\\
r_{s}=2M,\\
r_{Q}=Q^{2}
\end{split}
\end{equation}
Spacetime topology being $R\times R\times S^{2}$. The electromagnetic part is given as (in B-L coordinates)
\begin{equation}
A_{\mu}=(\frac{r r_{Q}}{\rho^{2}},0,0,-\frac{a r r_{Q}\sin^{2}\theta}{\rho^{2}M}) 
\end{equation}
The horizons are obtained by solving the equations 
$\Delta=0$
Solving it and we find the value of 
\begin{equation}
r_{0}=\frac{r_s\pm \sqrt{r_s^2-4(a^2+r_Q^2)}}{2} 
\end{equation}
 for the horizon.
Now let's write the Einstein equations 
\begin{equation}
R_{\mu\nu}-\frac{1}{2}Rg_{\mu\nu}=T_{\mu\nu}
\end{equation}
whose trace yields 
\begin{equation}
R=-T
\end{equation}
The Maxwell field has the stress tensor 
\begin{equation}
T^{\mu\nu}=F^{\mu\alpha}F_{\alpha}^{\nu}-\frac{1}{4}g^{\mu\nu}F^{2}\
\end{equation}
which clearly has trace $T=0$ and therefore $R=0$.
In 3+1 approach we use foliation of spacetime where the spacetime hypersurfaces are split in constant time surfaces. Yhe three dimension hypersurfaces have intrinsic curvature given by the Riemanian tensor while the hypersurface of constant time has extrinsic curvature given by the second fundamental form $K_{ij}$ arising from the embedding of the 3D hypersurface in the four dimensional spacetime.
 After implementing the Einstein equations, the partition function becomes 
\begin{equation}
I[g_{0},A_{0}]=-\frac{1}{4}\int_{M}d^{4}x\sqrt{-det(g_{\mu\nu})}
F_{\mu\nu}F^{\mu\nu}+\int_{\partial M}K d\partial M
\end{equation}

The action for the metric g over a region Y with boundary $\partial M$ is given by Gibbons and Hawking as 
\begin{equation}
I=(16\pi)^{-1}\int_{\partial M} R(-g)^{\frac{1}{2}}d^4x+\int_{d\partial M}B(-h)^{\frac{1}{2}}d^3x
\end{equation}
The surface term B is chosen such that it satisfies Einstein's equations and vanish at the boundary
\begin{equation}
B=(8\pi)^{-1}K
\end{equation}
where K is the trace of the second fundamental form at the boundary in metric $g$.
Now we focus on evaluating this background action functional explicitly in the Kerr-Newman spacetimes. Note the important features of this spacetimes. It has a time-Killing field $\frac{\partial}{\partial t}$ i.e., 
\begin{equation}
L_{\frac{\partial}{\partial t}}g_{0}=0,
\end{equation} 
$L$ being Lie derivative.  Here onwards for simplicity, $g_{0}$ is denoted by $g$. In addition, we have 
\begin{equation}
<\frac{\partial}{\partial r},\frac{\partial}{\partial t}>=<\frac{\partial}{\partial r},\frac{\partial}{\partial \theta}>=<\frac{\partial}{\partial r},\frac{\partial}{\partial \phi}>=0.\end{equation}
Clearly now, we may take $\partial M$ to be a constant $r$ hypersurface. The unit normal vector may be taken to be one which is proportional to $\frac{\partial}{\partial r}$ that is $n=a \frac{\partial}{\partial r}$ satisfies 
\begin{equation}
g(a\frac{\partial}{\partial r},a \frac{\partial}{\partial r})=-1
\end{equation}

We compute $a$ from this condition using the Kerr-Newman metric and thus the normal vector becomes 
\begin{equation}
n=\frac{\sqrt{\Delta}}{\rho} \frac{\partial}{\partial r}
\end{equation}
 The second fundamental form $K_{ij}$ of the boundary may be computed as follows 
\begin{equation}
K_{ij}=<\nabla_{\frac{\partial}{\partial x_{i}}}\frac{\partial}{\partial x_{j}},n>\\
=\Gamma^{\mu}_{ij}g_{\mu\lambda}n^{\lambda} 
\end{equation} 
here $i,j$ are $t$, $\theta$, and $\phi$. Using the expression of $n$ obtained previously,we can explicitly compute $K$. Now take the trace of $K$ with respect to $h_{ij}$ that is $h^{ij}K_{ij}$, which is the metric of the boundary that is after setting $r=r_{0}$ in the line element  (metric $h^{ij}$ may simply be found by inverting $h_{ij}$).
\begin{equation}
\mathsf{h_{ij}}=\left(
\begin{matrix} 
\frac{\Delta-a^2\sin^2\theta}{\rho^2} & 0 & \frac{2a\sin^2\theta(r^2+a^2-\Delta)}{\rho^2}\\
0 & -\rho^2 & 0\\
\frac{2a\sin^2\theta(r^2+a^2-\Delta)}{\rho^2} & 0 & \frac{\sin^2\theta}{\rho^2}[a^2\sin^2\theta \Delta- (r^2+a^2)^2]
\end{matrix}\right)
\end{equation}
 \\
 \\
 \begin{equation}
\mathsf{h^{ij}}=\left(
\begin{matrix} 
\frac{\rho^2((r^2+a^2)^2-a^2\sin^2\theta\Delta}{\Sigma} & 0 & -\frac{2a\rho^2(\Delta-1)}{\Sigma}\\
0 & -\frac{1}{\rho^2} & 0\\
-\frac{2a\rho^2(\Delta-1)}{\Sigma} & 0 & -\frac{\rho^2(\Delta-a^2\sin^2\theta)}{\Sigma\sin^2\theta}
\end{matrix}\right)
\end{equation}

where 
\begin{equation}
\Sigma=a^4\sin^4\theta\Delta+3a^2\sin^2\theta\Delta^2-8a^2(r^2+a^2)\sin^2\theta\Delta+3a^2(r^2+a^2)\sin^2\theta+(r^2+a^2)^2\Delta
\end{equation}
 But
\begin{equation}
\int Kd\partial M=\frac{\partial}{\partial n}\int d\partial M
\end{equation}

where $\frac{\partial}{\partial n}\int d\partial M$ is the derivative of the area $\int d\partial M$ of $\partial Y$ as each point of $\partial Y$ is moved an equal distance along the outward normal n. The regularity of the metric at the horizon requires that the point $(t,r,\theta,\phi)$ to be identified with $(t+\iota2\pi\kappa^{-1},r,\theta,\phi)$ where $\kappa$ is given by Hawking as 
\begin{equation}
\begin{split}
\kappa=(r_{+}-r_{-})2^{-1}(r_{+}^2+J^2M^{-2})^{-1}\\
r_{\pm}=(M\pm(M^2-J^2M^{-2}-Q^2)^{\frac{1}{2}}
\end{split}
\end{equation}
Thus, in Kerr-Newman solution near the ergosphere where $\Delta\rightarrow 0$
\begin{equation}
\begin{split}
\int Kd\partial M=\int n\sqrt{det(h_{\mu\nu})}=\frac{\sqrt{\Delta}}{\rho} \frac{\partial}{\partial r}\int\sqrt{-det(h_{\mu\nu})}\\
4\pi^3\iota\kappa^{-1}\frac{\sqrt{\Delta}}{\rho}[\frac{3\sqrt{3}ar(\rho-(r^2+a^2))-2\sqrt{3}a(\rho(2r-r_s)-2\Delta r)}{3\rho^2}\sin^2\theta\\
+\frac{\sqrt{3}a^3(\rho(r^2+a^2)(2r-r_s)-2\Delta r(\rho+r^2+a^2))}{16\rho^2(r^2+a^2)^2}\sin^4\theta\\
+\frac{\rho(\sqrt{3}(r^2+a^2)(2r-r_s)+2\sqrt{3}\Delta r)-2\sqrt{3}\Delta r(r^2+a^2)}{6a\rho^2}]
\end{split}
\end{equation}

Let us now concentrate on the Electromagnetic sector. The equation of motion is 
\begin{equation}
\begin{split}
\nabla_{\mu}F^{\mu\nu}=0,\\
\nabla^{\mu}\nabla_{\mu}A^{\nu}-\nabla_{\mu}\nabla^{\nu}A^{\mu}=0.
\end{split}
\end{equation} 
The quadratic term in the Lagrangian may be simplified as follows
\begin{equation}
\begin{split}
F_{\mu\nu}F^{\mu\nu}=(\nabla_{\mu}A_{\nu}-\nabla_{\nu}A_{\mu})(\nabla^{\mu}A^{\nu}-\nabla^{\nu}A^{\mu}),\\
=2(\nabla_{\mu}A_{\nu}\nabla^{\mu}A^{\nu}-\nabla_{\nu}A_{\mu}\nabla^{\mu}A^{\nu}),\\
=2\left\{\nabla_{\mu}(A_{\nu}\nabla^{\mu}A^{\nu})-A^{\nu}\nabla^{\mu}\nabla_{\mu}A_{\nu}-\nabla_{\mu}(A^{\nu}\nabla_{\nu}A^{\mu})+A^{\nu}\nabla_{\mu}\nabla_{\nu}A^{\mu}\right\},\\\nonumber
=2\left\{\nabla_{\mu}(A_{\nu}(\nabla^{\mu}A^{\nu}-\nabla^{\nu}A^{\mu}))-A_{\nu}(\nabla^{\mu}\nabla_{\mu}A^{\nu}-\nabla_{\mu}\nabla^{\nu}A^{\mu})\right\}.
\end{split}
\end{equation}
Using equation of motion, this reduces to 
\begin{equation}
F_{\mu\nu}F^{\mu\nu}=2\nabla_{\mu}(A_{\nu}(\nabla^{\mu}A^{\nu}-\nabla^{\nu}A^{\mu})).
\end{equation}
Therefore, the action reduces to 
\begin{equation}
I[g_{0},A_{0}]=-\frac{1}{2}\int_{\partial M}F_{\mu\nu}A^{\nu}n^{\mu}d\partial M+\int_{\partial M}K d\partial M.
\end{equation}

where $A^{\nu}$ is 
\begin{equation}
A^{\nu}=\left(\frac{rr_Q}{\Sigma}\left[(r^2+a^2)^2-a^2\sin^2\theta\Delta+\frac{2a^2(\Delta-1)}{M}\right],0,0,\frac{rr_Q}{\Sigma}\left[\frac{a(\Delta-a^2\sin^2\theta)}{M}-2a(\Delta-1)\right]\right)
\end{equation}
where $\Sigma$ is given by equation 2.18.

\begin{equation}
n^{\mu}=\left(0,\frac{\sqrt{\Delta}}{\rho}\frac{\partial}{\partial r},0,0\right)
\end{equation}

Thus, only two terms in $F_{\mu\nu}$ are important 
\begin{equation}
\begin{split}
F_{rt}=\nabla_r A_{t}=\frac{\partial A_t}{\partial r}=\frac{r_Q}{\rho^4}(\rho^2-2r^2)\\
F_{r\phi}=\nabla_r A_{\phi}=\frac{\partial A_{\phi}}{\partial r}=\frac{a\sin^2\theta}{M}(2r^2-\rho^2)
\end{split}
\end{equation}

Therefore,
\begin{equation}
\begin{split}
\int_{\partial M}F_{\mu\nu}A^{\nu}n^{\mu}d\partial M=\int\frac{rr_Q^2\sqrt{\Delta}}{\Sigma\rho^5}(\rho^2-2r^2)\left[(r^2+a^2)^2-a^2\sin^2\theta\Delta+\frac{2a^2(\Delta-1)}{M}\right]\\
\frac{\partial}{\partial r}\sqrt{-det(h_{\mu\nu})}dtd\theta d\phi\\+\int \frac{arr_Q\sqrt{\Delta}\sin^2\theta}{M\Sigma\rho}(2r^2-\rho^2)\left[\frac{a(\Delta-a^2\sin^2\theta)}{M}-2a(\Delta-1)\right]\\\frac{\partial}{\partial r}\sqrt{-det(h_{\mu\nu})}dtd\theta d\phi
\end{split}
\end{equation}

\section{The Action with Second order perturbation}
The partition function is given by
\begin{equation}
Z=\int[dg]e^{-\int_{M}d^{4}x\sqrt{-det(g_{\mu\nu})}(R(g)-\frac{1}{4}F_{\mu\nu}F^{\mu\nu})}.
\end{equation}
We can write the scalar curvature $R$ explicitly in terms of $g_{\mu\nu}$. This will contain the second derivatives which may be reduced to terms containing only boundary integral by integration by parts. This boundary integral may be related to the second fundamental form of the boundary already calculated in the previous section
\begin{equation}
K(u,v)=g_{\mu\nu}u^{\lambda}\nabla_{\lambda}v^{\nu}n^{\mu}.
\end{equation}
In order to cancel the boundary term, we need to add the negative of the boundary term to the action
\begin{equation}
Z=\int[dg][dA]e^{\iota\int_{M}d^{4}x\sqrt{-det(g_{\mu\nu})}(R(g)-\frac{1}{4}F_{\mu\nu}F^{\mu\nu})+\int_{\partial M}K d\partial M},
\end{equation}
where $d\partial M$ is the volume form on the boundary of the spacetime i.e., $\partial M$ and $K$ is the trace of the second fundamental form of the boundary with respect to the induced metric on the boundary. We will focus on the action term 
\begin{equation}
I[g,A]=\int_{M}d^{4}x\sqrt{-det(g_{\mu\nu})}(R(g)-\frac{1}{4}F_{\mu\nu}F^{\mu\nu})+\int_{\partial M}K d\partial M.
\end{equation}
The quantum partition function may be computed by the method of stationary phase after splitting the metric into background and perturbations
\begin{equation}
\begin{split}
g_{\mu\nu}=g_{0\mu\nu}+\delta g_{\mu\nu},\\
A_{\mu}=A_{0\mu}+\delta A_{\mu},
\end{split}
\end{equation}
where $f(A)$ is the gauge fixing term, which will eventually introduce ghosts terms. The action takes the form 
\begin{equation}
I[g,A]=I[g_{0},A_{0}]+I[\delta g,\delta A],
\end{equation}
which upon substituting into the partition unction yields 
\begin{equation}
\ln Z=\iota I[g_{0},A_{0}]+\ln \int[dg]e^{\iota I[\delta g,\delta A]}.
\end{equation}

The partition function with source term is given by 
\begin{equation}
 Z[J]=\int\mathcal{D}\phi e^{\frac{\iota}{\hbar}\int \mathcal{L} d^4x+\frac{\iota}{\hbar}J\phi}
\end{equation}

where $\mathcal{L}$ is the Lagrangian and $J$ is the source term.

For a free field the partition function becomes 
\begin{equation}
\begin{split}
Z[J]=\int\mathcal{D}\phi e^{\frac{\iota}{\hbar}\int d^4x[-\frac{1}{2}\phi(\square+m^2)\phi+\frac{1}{2}J\phi_0]}\\
=\int \mathcal{D}\phi e^{\frac{\iota}{\hbar}\int d^4x[-\frac{1}{2} \phi(\square+m^2)\phi]} e^{-\frac{\iota}{2\hbar}\int d^4xd^4yJ(x)\triangle_F(x-y)J(y)}\\
=Ne^{\iota W[J]}
\end{split}
\end{equation}
where $N$ is basically the Zeroth order term and $W[J]$ is the source term.

\begin{equation}
\begin{split}
W[J]=-\frac{1}{2\hbar}\int d^4xd^4yJ^i(x)D_{ij}(x-y)J^j(y)\\
=-\frac{1}{2\hbar} \frac{1}{(2\pi)^4} \int d^4k j^i(-k)D_{ij}(k)J^j(k)
\end{split}
\end{equation}
The Proca action describes a massive spin-1 field of mass m in Minkowski spacetime. The corresponding equation is a relativistic wave equation called the Proca equation. The Proca Lagrangian for an Electromagnetic field is given by 
\begin{equation}
\mathcal{L}=-\frac{1}{4} F_{ij}F^{ij}+J^iA_i
\end{equation}

where the first term is the electromagnetic term and the second term is the source term.
Calculating the electromagnetic term
\begin{equation}
-\frac{1}{4}F_{ij}F^{ij}=\frac{1}{2}(A_j\square g^{jk}A_k-A_j\nabla^k\nabla^jA_k)
\end{equation}

Thus the action becomes
\begin{equation}
I=\int d^4x [\frac{1}{2}A_j(g^{jk}\square-\nabla^k\nabla^j)A_k+J^iA_i]
\end{equation}

The zeroth order action is given by
\begin{equation}
\begin{split}
N=\int \mathcal{D}A e^{\frac{\iota}{2\hbar}\int d^4x(A_j(g^{jk}\square-\nabla^k\nabla^j)A_k)}\\
=\mathcal{D}Ae^{\frac{\iota}{2\hbar}\int \sqrt{-det(g_{\mu\nu})} dtd\theta dr d\phi [I]}
\end{split}
\end{equation}

where 
\begin{equation}
\begin{split}
[I]=A_j(g^{jk}\square-\nabla^k\nabla^j)A_k\\
=A_0(g^{00}(\nabla_r^2+\nabla_\theta^2))A_0+A_3(g^{33}(\nabla_r^2+\nabla_\theta^2))A_3\\
+A_0(g^{03}(\nabla_r^2+\nabla_\theta^2))A_3
+A_3(g^{30}(\nabla_r^2+\nabla_\theta^2))A_0
\end{split}
\end{equation}

Considering only the gravity term the action becomes

\begin{equation}
I[g]=\int_{M}d^{d}x\sqrt{det(g_{\mu\nu})}\left(R\right).
\end{equation}

Now we will apply perturbation to the curvature term to evaluate the action.

Let us just write down some important identities
\begin{equation}
\begin{split}
D\Gamma^{\mu}_{\nu\lambda}\cdot h=\frac{1}{2}g^{\mu\alpha}\left(\nabla_{\nu}h_{\alpha\lambda}+\nabla_{\lambda}h_{\nu\alpha}-\nabla_{\alpha}h_{\nu\lambda}\right),\\
DR_{\sigma\nu}\cdot h=\frac{1}{2}\left(\nabla_{\alpha}\nabla_{\sigma}h^{\alpha}_{\nu}+\nabla_{\alpha}\nabla_{\nu}h^{\alpha}_{\sigma}-\nabla^{\alpha}\nabla_{\alpha}h_{\sigma\nu}-\nabla_{\nu}\nabla_{\sigma}h^{\alpha}_{\alpha}\right)
\end{split}
\end{equation}
The total variation of $R_{\mu\nu}$ may be written as
\begin{equation}
R_{\mu\nu}(g)=R_{\mu\nu}(g_{0})+DR_{\mu\nu}\cdot h+\frac{1}{2}D^{2}R_{\mu\nu}\cdot (h,h)+......... ,
\end{equation}
and
\begin{equation}
R[g]=R[g_{0}]-R_{\mu\nu}h^{\mu\nu}+g_{0}^{\mu\nu}DR_{\mu\nu}\cdot h+\frac{1}{2}g_{0}^{\mu\nu}D^{2}R_{\mu\nu}\cdot (h,h)r -h^{\alpha\beta}DR_{\alpha\beta}\cdot h+ R_{\alpha\beta}(g_{0})h^{\alpha\mu}h^{\beta}_{\mu}....
\end{equation}
The second variation of $R_{\alpha\beta}$ may be computed as
\begin{equation}
\begin{split}
D^{2}R_{\alpha\beta}\cdot (h,h)=-h^{\lambda\mu}\left(\nabla_{\lambda}(\nabla_{\alpha}h_{\beta\mu}+\nabla_{\beta}h_{\alpha\mu}-\nabla_{\mu}h_{\alpha\beta})-\nabla_{\alpha}\nabla_{\beta}h_{\lambda\mu}\right),
\\-(\nabla_{\lambda}h^{\lambda\mu})(\nabla_{\alpha}h_{\beta\mu}+\nabla_{\beta}h_{\alpha\mu}-\nabla_{\mu}h_{\alpha\beta})+\frac{1}{2}(\nabla_{\beta}h^{\lambda\mu})(\nabla_{\alpha}h_{\lambda\mu})
\\+\frac{1}{2}(\nabla^{\lambda}h^{\rho}_{\rho})(\nabla_{\alpha}h_{\beta\lambda}+\nabla_{\beta}h_{\alpha\lambda}-\nabla_{\lambda}h_{\alpha\beta})+(\nabla_{\lambda}h^{\mu}_{\alpha})(\nabla^{\lambda}h_{\beta\mu}),
\\-(\nabla_{\lambda}h^{\mu}_{\alpha})(\nabla_{\mu}h^{\lambda}_{\beta}),
\end{split}
\end{equation}
\begin{equation}
\begin{split}
\frac{1}{2}D^{2}R\cdot(h,h)=\frac{1}{2}[-h^{\lambda\mu}\left(\nabla_{\lambda}(2\nabla_{\alpha}h^{\alpha}_{\mu}-\nabla_{\mu}h^{\alpha}_{\alpha})-\nabla^{\alpha}\nabla_{\alpha}h_{\lambda\mu}\right)
\\-(\nabla_{\lambda}h^{\lambda\mu})(2\nabla_{\alpha}h^{\alpha}_{\mu}-\nabla_{\mu}h^{\alpha}_{\alpha})+\frac{3}{2}(\nabla^{\alpha}h^{\lambda\mu})(\nabla_{\alpha}h_{\lambda\mu})
\\+\frac{1}{2}(\nabla^{\lambda}h^{\rho}_{\rho})(2\nabla^{\alpha}h_{\alpha\lambda}-\nabla_{\lambda}h^{\alpha}_{\alpha})-(\nabla_{\lambda}h_{\mu\alpha})(\nabla^{\mu}h^{\lambda\alpha}),
\\-h^{\alpha\beta}\nabla_{\mu}\nabla_{\alpha}h^{\mu}_{\beta}+\frac{1}{2}h^{\alpha\beta}\nabla^{\mu}\nabla_{\mu}h_{\alpha\beta}+\frac{1}{2}h^{\alpha\beta}\nabla_{\alpha}\nabla_{\beta}h^{\mu}_{\mu}
\\+R_{\alpha\beta}(g_{0})h^{\alpha\mu}h^{\beta}_{\mu}.
\end{split}
\end{equation}
\begin{equation}
\begin{split}
R(g)_{0}=R(g_{0}),
R(g)_{1}=g^{\alpha\beta}DR_{\alpha\beta}\cdot h-R_{\alpha\beta}h^{\alpha\beta},\\
=\nabla_{\mu}\nabla_{\alpha}h^{\mu\alpha}-\nabla^{\mu}\nabla_{\mu}h^{\alpha}_{\alpha}-R_{\alpha\beta}h^{\alpha\beta},\\
R(g)_{2}=\frac{1}{2}D^{2}R\cdot(h,h).
\end{split}
\end{equation}
Also we have the expansion
\begin{equation}
\mu_{g}=\mu_{g_{0}}(1+\frac{1}{2}h^{\mu}_{\mu}-\frac{1}{4}h^{\mu}_{\alpha}h^{\alpha}_{\mu}+\frac{1}{8}(h^{\mu}_{\mu})^{2}+....)
\end{equation}
where $\mu_{g}=\sqrt{-det(g)}$ is the volume element of the metric space.
Let us now write the Lagrangian with this perturbed entities
\begin{equation}
\begin{split}
I[g]=\int_{M}d^{d}x\mu_{g_{0}}(1+\frac{1}{2}h^{\mu}_{\mu}-\frac{1}{4}h^{\mu}_{\alpha}h^{\alpha}_{\mu}+\frac{1}{8}(h^{\mu}_{\mu})^{2})\\
(R(g_{0})+DR\cdot h+\frac{1}{2}D^{2}R\cdot(h,h)))\\
=\int_{M}d^{d}x\mu_{g_{0}}(1+\frac{1}{2}h^{\mu}_{\mu}-\frac{1}{4}h^{\mu}_{\alpha}h^{\alpha}_{\mu}+\frac{1}{8}(h^{\mu}_{\mu})^{2})\\
(R(g_{0})+DR\cdot h+\frac{1}{2}D^{2}R\cdot (h,h))
\end{split}
\end{equation}
By virtue of the field equations, the first order variation vanishes. The second order terms may be expressed fully in the following section. Let us first compute some unknown terms
\begin{equation}
\begin{split}
DR^{\alpha\beta}\cdot h=\frac{1}{2}\left(\nabla_{\lambda}\nabla^{\alpha}h^{\lambda\beta}+\nabla_{\lambda}\nabla^{\beta}h^{\lambda\alpha}-\nabla^{\lambda}\nabla_{\lambda}h^{\alpha\beta}-\nabla^{\alpha}\nabla^{\beta}h^{\lambda}_{\lambda}\right)-R^{\alpha}_{\nu}h^{\beta\nu}-R^{\beta}_{\nu}h^{\alpha\nu}\\
=\frac{1}{2}(\nabla^{\alpha}\nabla_{\lambda}h^{\lambda\beta}+\nabla^{\beta}\nabla_{\lambda}h^{\lambda\alpha}+R^{\alpha}_{\nu}h^{\nu\beta}+R^{\beta}_{\nu}h^{\nu\alpha}+R^{\beta}_{\nu\lambda}~^{\alpha}h^{\lambda\nu}+R^{\alpha}_{\nu\lambda}~^{\beta}h^{\lambda\nu}\\
-\nabla^{\lambda}\nabla_{\lambda}h^{\alpha\beta}-\nabla^{\alpha}\nabla^{\beta}h^{\lambda}_{\lambda})-R^{\alpha}_{\nu}h^{\beta\nu}-R^{\beta}_{\nu}h^{\alpha\nu}
\end{split}
\end{equation}

We will now explicitly write different quadratic terms
\begin{equation}
\begin{split}
(DR_{\alpha\beta}\cdot h)(DR^{\alpha\beta}\cdot h)=\frac{1}{4}(\nabla_{\alpha}\nabla_{\lambda}h^{\lambda}_{\beta}+\nabla_{\beta}\nabla_{\lambda}h^{\lambda}_{\alpha}+R_{\lambda\alpha}h^{\lambda}_{\beta}+R_{\lambda\beta}h^{\lambda}_{\alpha}\\
-R^{\lambda}_{\beta\mu\alpha}h^{\mu}_{\lambda}-R^{\lambda}_{\alpha\mu\beta}h^{\mu}_{\lambda}-\nabla^{\mu}\nabla_{\mu}h_{\alpha\beta}-\nabla_{\beta}\nabla_{\alpha}h^{\mu}_{\mu})\\
(\nabla^{\alpha}\nabla_{\lambda}h^{\lambda\beta}+\nabla^{\beta}\nabla_{\lambda}h^{\lambda\alpha}-R^{\alpha}_{\nu}h^{\nu\beta}-R^{\beta}_{\nu}h^{\nu\alpha}\\
+R^{\beta}_{\nu\lambda}~^{\alpha}h^{\lambda\nu}+R^{\alpha}_{\nu\lambda}~^{\beta}h^{\lambda\nu}-\nabla^{\lambda}\nabla_{\lambda}h^{\alpha\beta}-\nabla^{\alpha}\nabla^{\beta}h^{\lambda}_{\lambda}).
\end{split}
\end{equation}
The background manifold is not defined by any unique and natural
way, but by some non-trivial method of identifying all solution manifolds. In our case, we have introduced a particular geometric coordinate system in each solution manifold, and then identified those points of different solution manifolds that have
the same values of these coordinates. The choice of particular geometric coordinates in all solutions is usually a consequence of a gauge choice (or coordinate condition).Hence, it is the gauge fixing that defines points of a background manifold in general relativity.
Note the following identities which will be required for gauge fixing later.
\begin{equation}
1=\int \mathcal{D}\alpha~\delta(G(h^{(\alpha)}_{\mu\nu})) \det(\frac{\delta G(h^{(\alpha)}_{\mu\nu})}{\delta \alpha})
\end{equation}
An obvious choice of $G(h^{(\alpha)}_{\mu\nu})$ could be the harmonic gauge choice ($\mathcal{C}(x)\equiv 0$)
\begin{equation}
G(h^{(\alpha)})=\nabla^{\nu}(h^{(\alpha)}_{\mu\nu}-\frac{1}{2}(tr_{g})h g_{\mu\nu})-\mathcal{C}(x)
\end{equation}
General Relativity is a gauge theory in the sense that it is invariant under diffeomorphic transformations.Diffemorphism may be seen as a local (gauged) version of translations $\delta x^{\mu} \longrightarrow a^{\mu}(x)$. In order for the theory to be diffeomorphism invariant, a covariant derivative $\nabla$ must replace partial derivatives $\partial$ (a general, dynamic metric g tensor must replace Minkowski metric $\eta$ as well).
Note that the spacetime manifold is actually an equivalence class of pairs
(M, g), where two metrics are viewed as equivalent if one can be obtained
from the other through the action of the diffeomorphism group Diff(M). The
metric is an additional geometric structure that does not necessarily solve
any field equation.
In the path integration, we have
\begin{equation}
\begin{split}
Z=\int\mathcal{D}h_{\mu\nu}~e^{\iota I[h_{\mu\nu}]}\\
=\int_{\mathcal{D}(T^{*}M\otimes T^{*}M)/diff_{0}(M) }\mathcal{D}h_{\mu\nu}~e^{\iota I[h_{\mu\nu}]}\int \mathcal{D}\alpha~\delta(G(h^{(\alpha)}_{\mu\nu})) \det(\frac{\delta G(h^{(\alpha)}_{\mu\nu})}{\delta \alpha})
\end{split}
\end{equation}
From the gauge invariance of the second order Lagrangian, we have
\begin{equation}
h^{X}_{\mu\nu}=h_{\mu\nu}+(L_{X}g)_{\mu\nu}=h_{\mu\nu}+\nabla_{\mu}X_{\nu}+\nabla_{\nu}X_{\mu}
\end{equation}
and therefore,
\begin{equation}
\begin{split}
G(h^{X})=\nabla^{\nu}(h^{X}_{\mu\nu}-\frac{1}{2}(tr_{g}h^{X})h^{X}_{\mu\nu})\\
=\nabla^{\nu}(h_{\mu\nu}-\frac{1}{2}(tr_{g}h)h_{\mu\nu})+\nabla^{\mu}\nabla_{\mu}X^{\nu}+R^{\nu}_{\mu}X^{\mu}
\end{split}
\end{equation}
We may now write the functional determinant as follows
\begin{equation}
\begin{split}
\det(\frac{\delta G}{\delta X})=\det((\nabla^{\alpha}\nabla_{\alpha})\delta^{\mu}_{\nu}+R^{\mu}_{\nu})
\end{split}
\end{equation}
The gauge fixing yields the following quadratic term
\begin{equation}
\mathcal{L}_{2}=(-\frac{1}{4}\nabla_{\mu}h_{\alpha\beta}\nabla^{\mu}h^{\alpha\beta}-\frac{1}{2}\nabla_{\lambda}h^{\mu}_{\alpha}\nabla_{\mu}h^{\lambda\alpha}
-R_{\lambda\alpha}h^{\lambda}_{\alpha}+R^{\lambda}_{\alpha\mu\beta}h^{\mu}_{\lambda})
\end{equation}

The second variation of the scalar curvature is given as
\begin{equation}
\begin{split}
R_{2}(g)=\frac{1}{2}(-h^{\lambda\mu}[\nabla_{\lambda}(\nabla_{\alpha}h^{\alpha}_{\mu}+\nabla^{\alpha}h_{\alpha\mu}-\nabla_{\mu}h^{\alpha}_{\alpha})-\nabla^{\alpha}\nabla_{\alpha}h_{\lambda\mu}]\\
-(\nabla_{\lambda}h^{\lambda\mu})(\nabla_{\alpha}h^{\alpha}_{\mu}+\nabla^{\alpha}h_{\alpha\mu}-\nabla_{\mu}h^{\alpha}_{\alpha})+\frac{3}{2}(\nabla^{\alpha}h^{\lambda\mu})(\nabla_{\alpha}h_{\lambda\mu})\\
+\frac{1}{2}(\nabla^{\lambda}h^{\mu}_{\mu})(\nabla_{\alpha}h^{\alpha}_{\lambda}+\nabla^{\alpha}h_{\alpha\lambda}-\nabla_{\lambda}h^{\mu}_{\mu})-(\nabla_{\lambda}h^{\mu}_{\alpha})(\nabla_{\mu}h^{\lambda\alpha}))\\
-\frac{1}{2}h^{\alpha\beta}(2\nabla_{\alpha}\nabla_{\lambda}h^{\lambda}_{\beta}-\nabla_{\mu}\nabla^{\mu}h_{\alpha\beta}-\nabla_{\alpha}\nabla_{\beta}h^{\mu}_{\mu}+2R_{\lambda\alpha}h^{\lambda}_{\beta}-2R^{\lambda}_{\alpha\mu\beta}h^{\mu}_{\lambda})
\end{split}
\end{equation}
and after imposing the gauge invariance

\begin{equation}
\begin{split}
\mathcal{L}_{2}=(-\frac{1}{4}\nabla^{\mu}\nabla_{\mu}h_{\alpha\beta}h^{\alpha\beta}-\frac{1}{2}\nabla_{\lambda}h^{\mu}_{\alpha}\nabla_{\mu}h^{\lambda\alpha}+R^{\lambda}_{\alpha\mu\beta}h^{\mu}_{\lambda}h^{\alpha\beta})\\
=-\frac{1}{4}\left(h_{\alpha\beta}\nabla^{\mu}\nabla_{\mu}h^{\alpha\beta}+2h^{\lambda\alpha}\nabla^{\mu}\nabla_{\lambda}h_{\mu\alpha}-2R^{\lambda}_{\alpha\mu\beta}h^{\mu}_{\lambda}h^{\alpha\beta}\right)\\
+full~covariant~derivative~terms\\
=-\frac{1}{4}\left(h_{\alpha\beta}g^{\alpha\mu}g^{\beta\delta}\nabla^{2}h_{\mu\delta}+2h_{\alpha\beta}g^{\beta\delta}\nabla^{\mu}\nabla^{\alpha}h_{\mu\delta}-2h_{\alpha\beta}R^{\delta\alpha\mu\beta}h_{\mu\delta}\right)\\
=h_{\alpha\beta}G^{\alpha\beta\mu\delta}h_{\mu\delta}
\end{split}
\end{equation}
where
\begin{equation}
G^{\alpha\beta\mu\delta}=-\frac{1}{4}\left(g^{\alpha\mu}g^{\beta\delta}\nabla^{2}+2g^{\beta\delta}\nabla^{\mu}\nabla^{\alpha}-2R^{\delta\alpha\mu\beta}\right),
\end{equation}
and
\begin{equation}
\nabla^{4}=\nabla^{\mu}\nabla_{\mu}\nabla^{\alpha}\nabla_{\alpha},\\
\nabla^{2}=\nabla^{\mu}\nabla_{\mu}
\end{equation}

Now we do the wick rotation of time i.e., set $\tau=\iota t$. So equation (3.7) becomes 
\begin{equation}
\ln Z=- I[g_{0},A_{0}]+\ln \int[dg]e^{- I[\delta g,\delta A]}.
\end{equation} 
doing so we compactify the time direction that is the topology of the boundary becomes $S^{1}\times S^{2}$, which is compact. So the integration of $\partial M$ should yield a finite result. The argument of the imaginary time compactification may be argued as follows. The partition function may be written as 
\begin{equation}
Z=tr e^{-\beta H}=<\phi|e^{-\beta H}|\phi>,
\end{equation}  
where $\beta=\frac{1}{T}$, and $H$ is the Hamiltonian. This could eventually be transformed into the path integral language, where the time $t=i\beta$ and the trace implies that the initial and the final states are the same and therefore in terms of imaginary time propagation, there is a periodicity in the imaginary time. So, transform the time everywhere to the imaginary time $\tau$ and do the integration over the horizon in the limit given in Hawking's paper. 
The entropy may be computed as 
\begin{equation}
S=\ln Z+\beta<E>=\frac{\partial}{\partial \beta}(\beta \ln Z).
\end{equation}

\section{Conclusion}
In 1973, Stephen Hawking and Jacob Bekenstein showed that black holes should slowly radiate away energy, which poses a problem. From the no-hair theorem, one would expect the Hawking radiation to be completely independent of the material entering the black hole. Nevertheless, if the material entering the black hole were a pure quantum state, the transformation of that state into the mixed state of Hawking radiation would destroy information about the original quantum state. This violates Liouville's theorem and presents a physical paradox. In July 2004, Stephen Hawking published a paper presenting a theory that quantum perturbations of the event horizon could allow information to escape from a black hole, which would resolve the information paradox. His argument assumes the unitarity of the AdS/CFT correspondence which implies that an AdS black hole that is dual to a thermal conformal field theory.\\
Hawking has calculated the action and partition function for a Schwarzchild Black Hole. The solution for a Schwarzchild black hole is quite simple; here the calculations are done for a generalized Kerr-Newman black hole for a generalized metric with curved spacetime.\\
Here, path integral formalism is used to calculate the second order perturbation term and the zeroth order term for the action. The first order term is always zero due to the definition of least action.\\
However, there is much scope for improvement in further generalizing the calculations. Further work can be done to calculate the higher order perturbation terms in gravity to obtain a renormalizable theory for gravity.

\newpage
% inline
{\Large References}\\
\\
\\
G.H. Gibbons' and S.W. Hawking. Cosmological event horizons, thermodynamics, and particle creation\\
Gibbons, G. W.; Hawking, S. W.Action integrals and partition functions in quantum gravity\\ 
L. Dolan and R. Jackiw, Phys. Rev. D 9, 3320 (1974).\\
 L. D. Faddeev and V. N. Popov, Phys. Lett. 25B, 29 (1967).\\
  L. Smarr, Phys. Rev. Lett. 30, 71 (1973);30, 521(E) (1973).\\
Peskin, Schroeder: An Introduction to Quantum Field Theory\\
R. P. Feynman and Hibbs, Quantum Mechanics and Path Integrals (McGraw-Hill; New York, -1965\\
R. M. Wald, Commun. Math. Phys. 45, 9 (1975).\\
R. Penrose, Phys. Rev. Lett. 14, 57 (1965).\\
S.W. Hawking, Commun. Math. Phys. 43, 199 (1975).  \\
S. W. Hawking, Phys. Rev. D 14, 2460 (1976). \\
S. W. Hawking, Phys. Rev. D 13, 191 (1976).  \\
S. W. Hawking. and R. Penrose, Proc. R. Soc. London A314, 529 (1970).\\ S. W. Hawking and G. F. R. Ellis, The L,arge Scale Structure of Spacetime (Cambridge Univ. Press, Cambridge, England, 1973).

\end{document}